%% file: main.tex
\documentclass[sigconf,natbib=true]{acmart}

\AtBeginDocument{%
  }

\setcopyright{acmlicensed}
\copyrightyear{2025}
\acmYear{2025}
\setcopyright{cc}
\setcctype{by-nc}
\acmConference[SIGIR '25]{Proceedings of the 48th International ACM SIGIR Conference on Research and Development in Information Retrieval}{July 13--18, 2025}{Padua, Italy}
\acmBooktitle{Proceedings of the 48th International ACM SIGIR Conference on Research and Development in Information Retrieval (SIGIR '25), July 13--18, 2025, Padua, Italy}\acmDOI{10.1145/3726302.3730321}
\acmISBN{979-8-4007-1592-1/2025/07}

\usepackage{multirow}
\usepackage{graphicx}
\usepackage[english]{babel}
\usepackage{enumitem}
\usepackage{listings}
\usepackage{xcolor}
\usepackage{amsmath,mathrsfs}
\usepackage{tcolorbox}
\usepackage{enumitem} 
\usepackage{ifthen}
\usepackage{hyperref}
\hypersetup{
    colorlinks=true,
    linkcolor=blue,
    filecolor=magenta,      
    urlcolor=blue,
    }

\newcommand{\minSign}{\textbf{\textcolor{blue}{$\downarrow$}}}
\newcommand{\maxSign}{\textcolor{blue}{$\uparrow$}}
\newcommand{\maxHighlight}[1]{\textbf{#1}}

\newcommand{\minHighlight}[1]{\textit{#1}}

\newcommand{\update}[1]{\textcolor{black}{#1}}

\newcommand{\gpt}{\textit{gpt-4o}}
\newcommand{\llama}{\textit{llama-3.2-90b }}
\newcommand{\gemini}{\textit{gemini-1.5-pro }}
\newcommand{\gptfour}{\textit{gpt-4o }}
\newcommand{\queryV}{q_{\text{v}}}
\newcommand{\queryPZero}{q_{\text{p0}}}
\newcommand{\queryPOne}{q_{\text{p1}}}
\newcommand{\Query}{\mathbb{Q}}

\newcommand{\KB}{\mathit{KB}}
\newcommand{\persona}{\mathit{p}}
\newcommand{\context}{\mathscr{D}}
\newcommand{\filter}{\mathit{f}}
\newcommand{\filters}{\mathcal{F}}
\newcommand{\QueryGenLLM}{\mathcal{L}_{\text{query\_gen}}}
\newcommand{\RecGenLLM}{\mathcal{L}_{\text{rec\_gen}}}
\newcommand{\recGenCities}{c_{rec\_gen}}

\newcommand{\link}[1]{\underline{#1}}

\newboolean{Yasharversion}
\setboolean{Yasharversion}{true}  %

\addto\extrasenglish{%
}

\begin{document}

\title[SynthTRIPs]{SynthTRIPs: A Knowledge-Grounded Framework for Benchmark Query Generation for Personalized Tourism Recommenders
}

\author{Ashmi Banerjee}
\email{ashmi.banerjee@tum.de}
\affiliation{%
  \institution{Technical University of Munich}
  \city{Munich}
  \country{Germany}
}
\author{Adithi Satish}
\email{adithi.satish@tum.de}
\affiliation{%
  \institution{Technical University of Munich}
  \city{Munich}
  \country{Germany}
}

\author{Fitri Nur Aisyah}
\email{fitri.aisyah@tum.de}
\affiliation{%
  \institution{Technical University of Munich}
  \city{Munich}
  \country{Germany}
}

\author{Wolfgang W\"orndl}
\email{woerndl@in.tum.de}
\affiliation{%
  \institution{Technical University of Munich}
  \city{Munich}
  \country{Germany}
}

\author{Yashar Deldjoo}
\email{yashar.deldjoo@poliba.it}
\affiliation{
  \institution{Polytechnic University of Bari}
  \city{Bari} 
  \country{Italy} 
}

\renewcommand{\shortauthors}{Banerjee et al.}

\begin{abstract}
  Tourism Recommender Systems (TRS) are crucial in personalizing travel experiences by tailoring recommendations to users’ preferences, constraints, and contextual factors. However, publicly available travel datasets often lack sufficient breadth and depth, limiting their ability to support advanced personalization strategies --- particularly for sustainable travel and off-peak tourism. 
  In this work, we explore using Large Language Models (LLMs) to generate synthetic travel queries that emulate diverse user personas and incorporate structured filters such as budget constraints and sustainability preferences.

  This paper introduces a novel SynthTRIPs framework for generating synthetic travel queries using LLMs grounded in a curated knowledge base (KB). Our approach combines persona-based preferences (e.g., budget, travel style) with explicit sustainability filters (e.g., walkability, air quality) to produce realistic and diverse queries. 
  We mitigate hallucination and ensure factual correctness by grounding the LLM responses in the KB. 
  We formalize the query generation process and introduce evaluation metrics for assessing realism and alignment. Both human expert evaluations and automatic LLM-based assessments demonstrate the effectiveness of our synthetic dataset in capturing complex personalization aspects underrepresented in existing datasets.
  While our framework was developed and tested for personalized city trip recommendations, the methodology applies to other recommender system domains.

  Code and dataset are made public at: \href{https://bit.ly/synthTRIPs}{\link{https://bit.ly/synthTRIPs}}

\end{abstract}



\maketitle

\input{sections/0_intro.tex}
\input{sections/2_related.tex}

\input{sections/3_SynthTRIPs}

\input{sections/4_Resources}

\input{sections/eval/0_main.tex}

\input{sections/recllm_eval.tex}

\input{sections/conclusion.tex}

\begin{acks}
We thank the Google AI/ML Developer Programs team for supporting us with Google Cloud Credits.
\end{acks}

\bibliographystyle{ACM-Reference-Format}
\bibliography{main}

\appendix

\end{document}

%% file: sections/0_intro.tex
\section{Introduction}

Recommender Systems have become indispensable in helping users sift through massive amounts of information in domains such as e‐commerce, media streaming, and travel~\cite{isinkaye2015recommendation}. In the travel domain, personalization is especially valuable: user choices depend on individual preferences (e.g., budget, travel style, sustainability concerns) and contextual factors (destination popularity, local activities, etc.)~\cite{balakrishnan2021multistakeholder}. Consequently, \textbf{high‐quality training data} that reflects real‐world travel preferences is essential for building effective and robust Tourism Recommender Systems (TRS).

However, publicly available travel datasets are often limited in \textbf{breadth} (e.g., focusing only on popular cities) and \textbf{depth} (e.g., missing nuanced preferences like climate consciousness, walkability, or budget constraints). The lack of richly annotated data hinders research on advanced personalization strategies incorporating specialized filters --- such as sustainable travel or off‐peak tourism.

    \emph{Large Language Models (LLMs)} --- such as GPT-4, PaLM, and LLaMA --- provide a powerful way to generate human-like text. They have performed remarkably in question-answering, summarization, and creative writing. In the travel context, LLMs can:
    \begin{enumerate}
        \item Emulate diverse \textit{user personas} (e.g., a budget-conscious student vs. a sustainability-focused family).
        \item Incorporate \textit{structured filters} (e.g., budget limits and environmental impact considerations).
        \item Generate \textit{natural-sounding requests} that integrate factual knowledge (e.g., city highlights, monthly visitor footfall). 
    \end{enumerate}

However, LLMs are prone to \emph{hallucination} when asked for domain-specific details. 
To mitigate this, we propose \emph{grounding} LLM responses in curated knowledge bases (KB) to ensure factual correctness regarding destinations, costs, or sustainability metrics.

Given that travel data with rich user and contextual features is notoriously difficult, we aim to share the insights as a \emph{resource paper}. In particular, our goal is to share\textit{ (i)} a synthetic dataset that captures realistic travel constraints; \textit{(ii)} a generation pipeline that enables flexible updates (new personas, destinations, or sustainability constraints); and \textit{(iii)} reproducible evaluation protocols. We hope this will help to address the pressing community need for better coverage, deeper personalization, and a more robust domain grounding in tourism recommendations.

\begin{figure}[!h] 
    \includegraphics[width=\columnwidth]{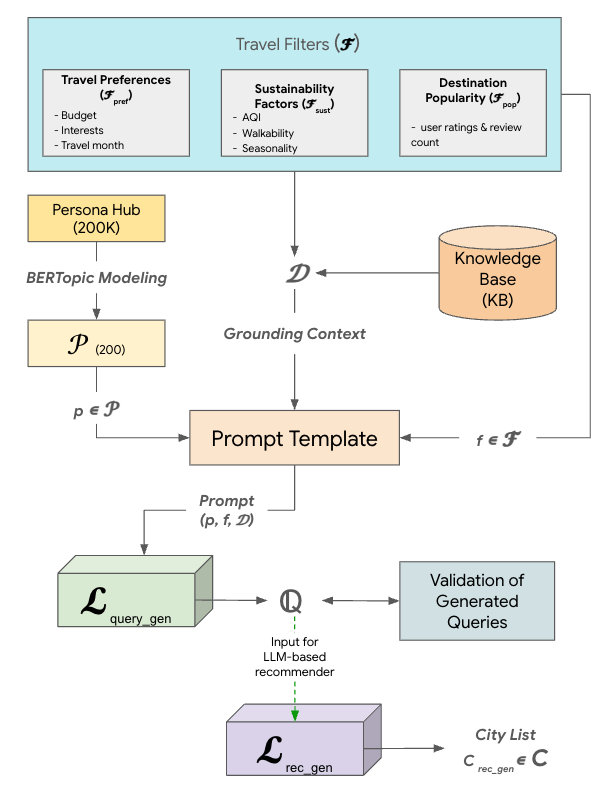}
    \caption{Our proposed framework for generating synthetic data using LLMs for personalized, sustainable city trips.}
    \label{fig: prompt template}
\end{figure}   

Below are some example user queries for a sustainable city trip, ensuring factual accuracy by grounding in a verified KB.  

    \begin{itemize}[leftmargin=15pt, itemsep=3pt] %
        \item \textbf{Q1.} \textit{``Find a low-budget, walkable city in Europe with unusual museums or a hidden, alternative nightlife scene.''}
        
        \item \textbf{Q2.} \textit{``Quiet European coastal city with good air quality, affordable, not touristy, with interesting nightlife options.''}
        
        \item \textbf{Q3.} \textit{``Best European cities for unique, artistic experiences and independent cinema, avoiding mainstream tourist attractions?''}
    \end{itemize}

Such queries capture complex \textbf{personalization} aspects (budget, interests, sustainability) that are typically underrepresented in publicly available travel datasets~\cite{xie2024travelplanner, wen_elaborative_2024}. 
\textit{As evident from~\autoref{fig: prompt template}, our resource generation framework, SynthTRIPs, automates the creation of such queries at scale, pairing persona definitions with domain-specific constraints to produce a rich corpus for training and benchmarking TRS models.}
    
\subsection*{Proposed Resource and Contributions}

This paper presents \textbf{SynthTRIPs}, a novel resource for building sustainable, persona-aware travel data via LLMs. SynthTRIPs addresses a critical gap in the tourism recommendation community by offering:

\begin{itemize}[leftmargin=15pt]
    \item \textbf{A reproducible pipeline for large-scale synthetic data generation} rooted in a curated KB that ensures factual grounding. By focusing on European city trips and integrating sustainability metrics such as walkability, off-peak travel, and popularity constraints, SynthTRIPs provides a versatile blueprint adaptable to diverse travel contexts.
    \item \textbf{Open-source datasets and code}: We publicly release all components, including:
    \begin{itemize}
        \item \textbf{Data}: 
        (1) A structured KB containing real-world attributes of European cities, 
        (2) Generated queries that integrate filters (budget, timing, popularity, sustainability)
        \item \textbf{Code for Query Generation}:
        (1) Prompt templates leveraging In-Context Learning (ICL) with various persona definitions,
        (2) Colab notebooks for straightforward replication of the entire pipeline.
        \item \textbf{Code for Evaluation}:
        (1) Automated LLM-based evaluation measuring groundedness, factual correctness, persona alignment, and sustainability compliance,
        (2) Tools for expert (human) evaluation for clarity, diversity, and overall rating.
    \end{itemize}
    \item \textbf{A flexible framework for future expansions}: Researchers can easily adapt SynthTRIPs to other regions or thematic focuses (beyond sustainability) by updating the underlying KB and modifying persona definitions or filter sets.
\end{itemize}

By offering a complete methodology for generating data and robust evaluation metrics, SynthTRIPs opens new avenues for more \textbf{personalized, sustainability-focused} travel recommendations. 

Our paper is organized as follows: \autoref{section: related} examines existing gaps in the literature and relevant work, while \autoref{section: SynthTRIPs-system} provides a detailed formal description and methodology of our query generation framework. \autoref{section: resources} subsequently presents our primary resources along with their key statistics. Following that, \autoref{section: validation} outlines the various validation dimensions utilized to evaluate the quality of our generated queries. Finally, \autoref{section: conclusion} concludes the paper, discussing its limitations and suggestions for future research.

%% file: sections/2_related.tex
\section{Related Work} \label{section: related}

    Early travel recommenders primarily relied on historical booking logs and user reviews \cite{lu2012personalized,lim2018personalized,chaudhari2020comprehensive}. While these sources offer insights into popular destinations, they rarely include explicit data for \emph{sustainability preferences} or off-peak travel. Subsequent works have integrated \emph{knowledge graphs} to enrich semantic understanding of destination attributes \cite{lan_knowledge_2022}, yet they often lack \emph{diverse user persona data}. A popular and effective way of considering the nuances of user preferences is through Conversational Recommender Systems, which provide more interactive recommendations compared to traditional recommenders~\cite{jannach2022conversational}. ~\autoref{tab:data_comparison} provides an overview of some of the datasets in this field, across domains like travel~\citep{meyer2024comparison,xie2024travelplanner,wen_elaborative_2024} and food~\cite{zhang_recipe-mpr_2023}. However, the dearth of data to evaluate this system is a persistent problem, especially with the time-consuming and resource-intensive nature of manual data collection and annotation.

    \begin{table*}
        \caption{An overview of the existing datasets and benchmarks as well as SynthTRIPs.}
        \label{tab:data_comparison}
        \centering 
        \resizebox{\textwidth}{!}{
        \begin{tabular}{@{}p{2.5cm} p{1.2cm} p{3cm} p{4.5cm} p{1.5cm} p{2.0cm} p{3.5cm} p{1cm}@{}}
        \toprule
        \multirow{2}{*}{\textbf{Dataset}}  & 
        \multirow{2}{*}{\textbf{Domain}} & 
        \multirow{2}{*}{\textbf{Goal/Task}} & 
        \multirow{2}{*}{\textbf{Description}}
        & \textbf{Size (\#queries)} & \textbf{LLMs/Model Used} & 
        \multirow{2}{*}{\textbf{Data Sources}} \\ \midrule
        reddit-travel-QA-finetuning~\cite{meyer2024comparison} & Travel & Question Answering + Finetuning & Real-time travel discussions (posts and top comments) on top 100 travel-related subreddits & 10,000 & N/A & Reddit \\
        TravelPlanner~\cite{xie2024travelplanner} & Travel & Travel Plan Generation & Synthetically generated queries and travel itineraries using LLMs, including constraints such as cuisine, accommodation, and transport & 1,225 & GPT-4 & Scraped from the Internet \\
        TravelDest~\cite{wen_elaborative_2024} & Travel & City Trip Recommendations & Broad and indirect queries spanning over 700 cities & 50 & Manually Generated & Wikivoyage \\
        FeB4RAG~\cite{wang2024feb4rag} & General & Federated Search & Information requests collected from the BEIR subcollections across multiple domains & 790 & GPT-4 & BEIR Datasets \cite{thakur2021beir} \\ 
        Recipe-MPR~\cite{zhang_recipe-mpr_2023} & Food & Multi-Aspect Recommendation & Manually annotated preference-based queries that contain multi-item descriptions to benchmark conversational recommender systems & 500 & Manually Generated & FoodKG~\cite{haussmann2019foodkg}, Recipe1M+~\cite{marin2021recipe1m+}\\ 
        \midrule
        \textbf{SynthTRIPs} & \textbf{Travel} & \textbf{City Trip Recommendations} & \textbf{Personalized travel queries  generated synthetically covering travel filters such as budget, interests integrated with additional sustainability features} & \textbf{4,604} & \textbf{Llama-3.2-90B, Gemini-1.5-Pro} & \textbf{Wikivoyage, Nomadlist, Where And When, Tripadvisor} \\ \bottomrule
        \end{tabular}
        }
    \end{table*}

    Meanwhile, with the recent advancements in LLMs and their ability to generate human-like conversations, there is a growing field of study that explores their potential in simulating users and their behavior~\citep{jandaghi2023faithful,zhu2024llm,abbasiantaeb2024let, wang2023improving, ashby2024towards, lu2023august, leszczynski2023talk}. LLM-based synthetic data generation has gained traction in domains such as healthcare \cite{tang2023does} and education \cite{khalil2025creating}, demonstrating the potential to automatically produce large-scale, high-quality text data.

    Nonetheless, existing approaches either (1) do not specifically tailor synthetic data to \emph{travel} use cases, like FeB4RAG~\cite{wang2024feb4rag}, or (2) do not incorporate robust \emph{sustainability and personalization} constraints. Furthermore, naive LLM-based query generation may lead to factual errors unless carefully grounded.

    Some studies have used LLMs to generate data for TRS, but they are often broad queries that do not necessarily involve a user asking for city trip recommendations. For example, the Reddit Q\&A dataset introduced by ~\citet{meyer2024comparison} consists of 10,000 posts and comments taken from various travel-related subreddits, covers a broad spectrum of travel-related queries, including but not limited to visas, required vaccines, and itinerary suggestions. TravelDest, a benchmark introduced~\cite{wen_elaborative_2024}, consists of 50 broad and indirect textual queries, which are insufficient to train algorithms. 
    In tourism,~\citet{xie2024travelplanner} uses LLMs to generate synthetic travel queries for origins and destinations in the United States to annotate and benchmark travel itineraries. 
    However, TravelPlanner~\cite{xie2024travelplanner} only focuses on testing the planning capabilities, so it doesn't consider the persona aspect of the query. All the queries sound similar and are in the same template, so we can't test preference extraction capabilities on this dataset.

    In contrast, our work directly addresses these \textbf{gaps} by:
    \begin{itemize}
        \item \textbf{Combining} persona-based preferences (budget, special interests) with explicit \emph{sustainability filters}.
        \item \textbf{Grounding} each query in a KB to reduce hallucination.
        \item \textbf{Introducing} a formal framework and evaluation metrics for query realism and alignment.
    \end{itemize}

%% file: sections/3_SynthTRIPs.tex
\section{SynthTRIPs: Query Generation System}
\label{section: SynthTRIPs-system}

We now describe the formal design of SynthTRIPs and illustrate how LLMs are leveraged to generate rich, realistic user queries. Our system is organized around three core modules: 
\textit{(1) Persona Hub}, \textit{(2) Travel Filters}, and \textit{(3) Contextual Prompting with KB Grounding}, as shown in Figure~\ref{fig: prompt template}. 

\subsection{Formal Description}
\label{subsection: formal description}

\paragraph{Personas:}

Prior research has shown that persona-guided LLMs can generate more diverse annotations than standard LLMs, improving the variability of LLM-based data annotation~\cite{frohling2024personas, ge2024scaling}.
Hence, to include personalization in our setup, we use personas from the PersonaHub\footnote{https://huggingface.co/datasets/proj-persona/PersonaHub} dataset, containing 200,000 unique personas.
This dataset is a repository of diverse user profiles (e.g., student, family, digital nomad), where each profile includes attributes such as age, interests, and environmental concerns.

To ensure broader coverage and reduce redundancy, we apply \textbf{BERTopic modeling}~\cite{grootendorst2022bertopic} using \texttt{all-MiniLM-L6-v2} embeddings on 196,693 English personas, clustering them into 201 topics (with one outlier group). We then select the highest probability persona from each cluster, resulting in a refined set of 200 diverse and representative personas.

Let \( \persona \in \mathcal{P} \), where \( \mathcal{P} = \{\persona_1, \persona_2, \dots, \persona_{n}\} \) where $n = 200$, denotes this refined set of personas, and let \( \filter \in \filters \) be a set of travel filters (explained below).

\paragraph{Knowledge Base (KB)}
We denote our knowledge base of European cities and their attributes as:
\[
\KB = (C, A, V),
\]
Where:
\begin{itemize}[leftmargin=15pt]
    \item $C$ is the set of \emph{city entities} (e.g., ``Amsterdam,'' ``Berlin,'' ``Budapest''),
    \item $A$ is the set of \emph{attribute types} relevant for city travel (e.g., ``walkability index,'' ``average cost,'' ``peak season,'' ``AQI''),
    \item $V: C \times A \to \mathcal{D}$ is a partial function mapping city-attribute pairs to a value domain $\mathcal{D}$ (e.g., real numbers or categorical labels).
\end{itemize}
We refer to $\KB$ as containing all factual knowledge used to ground the LLM generations.

\paragraph{Travel Filters:}
These filters $\filters$ are used to retrieve the relevant domain information from $\KB$ using structured queries. 
We define the set of travel filters \(\filters\) as a combination of \textbf{travel filters}, \textbf{sustainability factors}, and \textbf{destination popularity}:

\begin{equation}
    \filters = \filters_{\text{pref}} \cup \filters_{\text{sust}} \cup \filters_{\text{pop}}
\end{equation}

\noindent
\textbf{Travel filters} (\(\filters_{\text{pref}}\)):
\begin{equation}
    \filters_{\text{pref}} = \{ \filters_\text{budget}, \filters_\text{month}, \filters_\text{interests} \}
\end{equation}
Where:
\begin{itemize}[leftmargin=15pt]
    \item $\filters_\text{budget}$: (low, medium, high),
    \item $\filters_\text{month}$: (January--December),
    \item $\filters_\text{interests}$: from five categories~\cite{sanchez2022travelers}: 
    \textit{Arts \& Entertainment}, \textit{Outdoors \& Recreation}, \textit{Food}, \textit{Nightlife Spots}, \textit{Shops \& Services}.
\end{itemize}

\noindent
\textbf{Sustainability Factors} (\(\filters_{\text{sust}}\)):
\begin{equation}
    \filters_{\text{sust}} = \{ \filters_{\text{seasonality}}, \filters_{\text{walkability}}, \filters_{\text{AQI}} \}
\end{equation}
Here, $\filters_{\text{seasonality}}$ focuses on less crowded (off-peak) months, $\filters_{\text{walkability}}$ imposes a minimum walkability index, and $\filters_{\text{AQI}}$ checks city air quality thresholds. Sustainable destinations typically prioritize low seasonality, high walkability, and excellent AQI.

\noindent
\textbf{Destination Popularity} (\(\filters_{\text{pop}}\)):
We estimate popularity from the normalized number of Tripadvisor reviews~\cite{banerjee2025modeling}. For a city $c$, let $R(c)$ be the number of reviews. We define:
\begin{equation}
    \mathrm{popularity}(c) = \frac{R(c) - \min\limits_{c' \in C} R(c')}{\max\limits_{c' \in C} R(c') - \min\limits_{c' \in C} R(c')}
\end{equation}
We then split the normalized scores into three tiers (\textit{low}, \textit{medium}, \textit{high}). 
We separate popularity from sustainability constraints to ensure a balanced dataset covering all popularity ranges.

\paragraph{Complexity Levels of Filters.}
Inspired by TravelPlanner~\cite{xie2024travelplanner}, we define four levels of complexity:
\begin{align*}
    \textbf{Easy:} & \quad \filter_{\text{easy}} = \{ f \}, \quad f \in_R \filters_{\text{pref}} \\
    \textbf{Medium:} & \quad \filter_{\text{med}} = \{ f_1, f_2 \}, \quad f_1, f_2 \in_R \filters_{\text{pref}}, f_1 \neq f_2 \\
    \textbf{Hard:} & \quad \filter_{\text{hard}} = \filters_{\text{pref}} \\
    \textbf{Sustainable:} & \quad \filter_{\text{sust}} = \{ f_1, f_2 \} \cup \{ s \}, \quad f_1, f_2 \in_R \filters_{\text{pref}}, s \in_R \filters_{\text{sust}}
\end{align*}
This ensures a mix of simple and more complex queries. 
Formally, let 
\[
\filters_{\text{complexity}} = \{\filter_{\text{easy}}, \filter_{\text{med}}, \filter_{\text{hard}}, \filter_{\text{sust}}\}.
\]

\paragraph{Overall Key Functions.}
For each persona \(\persona \in \mathcal{P}\), we generate queries across all four complexity levels and three city popularity tiers (\textit{low}, \textit{medium}, \textit{high}). This yields:
\[
|\mathcal{P}| \times |\filters_{\text{complexity}}| \times |\filters_{\text{pop}}| = 200 \times 4 \times 3 = 2{,}400
\]
possible key functions, denoted \(\ell(\persona, \filter)\). 

\paragraph{Retrieval of Grounding Context.}
When a key function \(\ell(\persona, \filter)\) is chosen, we run a structured query on $\KB$:
\[
\context \subset \mathcal{D} \;=\; \mathrm{Retrieve}(\KB,\;\filter),
\]
Thus obtaining only the relevant facts (e.g., cities $c_\context$ and their corresponding attributes satisfying the filter constraints). If $\mathrm{Retrieve}(\KB, \filter) = \emptyset$, the filter is deemed invalid.
This results in 2302 valid key functions for our use case.

\paragraph{Prompt Construction.}
The context $\context$, along with the persona $\persona$ and the filter set $\filter$, is assembled into a structured prompt:
\[
\mathrm{Prompt}~(\persona, \filter, \context).
\]
We feed this prompt to a \emph{Query Generator LLM}, denoted \(\QueryGenLLM\). The resulting final user query is:
\[
\Query = \QueryGenLLM~\!\big(\mathrm{Prompt}(\persona, \filter, \context)\big).
\]

\paragraph{City Recommendation (Optional).}
If a subsequent step uses an LLM or any recommender model \(\RecGenLLM\) for city recommendations (given the newly generated user query), we denote that:
\[
\recGenCities \;=\; \RecGenLLM~(\Query),
\]

Where $\recGenCities$ is the list of recommended cities provided by $\RecGenLLM$. However, this step is optional for creating our \emph{synthetic user query} dataset since we primarily focus on generating queries in \textbf{SynthTRIPs}. Still, the framework naturally extends to a multi-turn pipeline.

\subsection{Goal} \label{section: goal}
The goal of SynthTRIPs is to produce a dataset $\Query = \{q_1, q_2, \dots\}$ and accompanying cities in the context $c_\context$, ensuring the queries are realistic, personalized, and grounded in factual knowledge.

\subsection*{Example} Consider a configuration where:
\begin{itemize}
    \item \textbf{Persona ($p$)}: \emph{"A wanderlust-filled trader who appreciates and sells the artisan's creations in different corners of the world"} 
    \item \textbf{Filters ($f$)}: 
    \begin{itemize}
        \item Popularity: Low
        \item Interests: Nightlife Spot
    \end{itemize}
\end{itemize}

Given this configuration, the pipeline generates a variety of queries:
\begin{itemize}
    \item \textbf{Vanilla Query ($q_{\text{v}}$)}: 
    \textit{"Recommend off-the-beaten-path European cities with low popularity for a nightlife-focused trip with a mix of bars, clubs, and live music venues."}
    \item \textbf{Persona-Specific Query 1 ($q_{\text{p0}}$)}: 
    \textit{"Unique nightlife and cultural experiences in off-the-beaten-path European cities for a budget-conscious traveler interested in local artisans."}
    \item \textbf{Persona-Specific Query 2 ($q_{\text{p1}}$)}: 
    \textit{"Which European cities offer a rich cultural heritage, historic centers, and local artisan markets to explore?"}
\end{itemize}

This example highlights how combining personas and filters shapes the generated queries, providing general (vanilla) and tailored (persona-specific) outputs.
We explain the different outputs (vanilla and personalized) in~\autoref{section: Response Gen}.

\subsection{In-Context Learning and Prompt Templates}
SynthTRIPs employs few-shot or single-shot \textbf{ICL} to guide the LLM to produce queries that are \emph{stylistically aligned} with the persona and \emph{faithful} to the filters. Our prompt template~\cite{deldjoo2023fairness} typically includes:
\begin{enumerate}[leftmargin=15pt]
    \item \textbf{Task Description}: Instruct the LLM to generate user queries given a specific persona, set of travel filters, and retrieved KB data.
    \item \textbf{In-Context Examples}: One or more example queries illustrating the desired style and structure.
    \item \textbf{Persona, Filters, \& Context}: The explicit persona traits, the chosen filter constraints, and relevant city attributes from the KB.
    \item \textbf{Output Requirements}: The final request to produce a single-sentence or multi-sentence user query, possibly with formatting constraints (e.g., no direct city names).
\end{enumerate}

\subsection{Response Generation} \label{section: Response Gen}

For each of the 2,302 valid key functions, we experiment with three conditions using \llama and \gemini as the $\QueryGenLLM$ with a \textit{temperature} setting of 0.5 and \textit{top\_p} value of 0.95:
\begin{itemize}
    \item \textbf{$\queryV$: Vanilla (Non-Personalized)}: Persona information is omitted.
    \item \textbf{$\queryPZero$: Personalized Zero-Shot}: Persona information is included, but without an example.
    \item \textbf{$\queryPOne$: Personalized Single-Shot}: Persona information is included along with an example.
\end{itemize}

\subsection*{Hallucination Mitigation via Grounding}

LLMs can produce \emph{hallucinated} statements when lacking access to factual references~\cite{huang2024survey}. To address this in SynthTRIPs, we explicitly \textbf{ground} the LLM output by:

\begin{itemize}[leftmargin=15pt]
    \item \textbf{Strict retrieval from $\KB$}: Only validated attributes from $\KB$ that meet the filter constraints are provided in the prompt.
    \item \textbf{Post-generation checks}: We implement an output parser to ensure properly formatted outputs. When the LLM generates overly verbose responses, we first apply regex-based extraction. If unsuccessful, we use an LLM-based parser using \gemini followed by manual verification.
\end{itemize}

SynthTRIPs significantly reduces inaccuracies and fosters factual consistency through prompt guidance and post-generation verifications.

%% file: sections/4_Resources.tex
\section{Resources} \label{section: resources}

This section outlines our key resources: the Knowledge Base (\autoref{subsection: KB}) used for query generation and the final dataset (\autoref{subsection: GenQueries}, which includes the generated queries along with their associated information.

\subsection{Knowledge Base} \label{subsection: KB}

Our $\KB$ construction is inspired by the data sources and methodology outlined in~\citet{banerjee2024enhancing}.
We utilize Wikivoyage's \textit{Listings}\footnote{https://github.com/baturin/wikivoyage-listings} data for detailed tabular information on POIs, including places to visit, accommodations, and dining options in 141 unique European cities.
The POIs in this dataset are categorized into different activity categories such as \textit{'see'}, \textit{'do'}, \textit{'eat'}, \textit{'drink'}, \textit{'sleep'}, and \textit{'go'}.

To map the wikivoyage listings POIs to the five travel preferences in~\autoref{subsection: formal description}, we use \texttt{bart-large-mnli}, a BART~\cite{lewis2019bart} variant trained on the MNLI dataset~\cite{williams2018mnli}, for zero-shot classification. This pre-trained Natural Language Inference (NLI) model enables sequence classification without additional training~\cite{yin-etal-2019-benchmarking} and assigns a probability score indicating how well each POI aligns with the defined travel filters.
The top three most probable items in each activity sub-category per city are selected to maintain consistency, reduce anomalies, and remain within the LLM's context window.

Publicly available data from Tripadvisor\footnote{\url{https://tripadvisor.com}} and Where and When (W2W)\footnote{\url{https://www.whereandwhen.net}} is utilized to incorporate both popularity and seasonality into the sustainability features of our $\KB$. Tripadvisor reviews and opinions represent city popularity, while W2W's monthly visitor footfall data indicates seasonality~\cite{banerjee2024enhancing}. 
Additional cost labels, AQI levels, and walkability data are sourced from Nomadlist\footnote{\url{https://nomadlist.com}}.

\subsubsection*{Basic Statistics}
The final dataset comprises 200 cities from 43 European countries, formed by combining the cities from the aforementioned four datasets.
Analyzing the distribution of popularity in our $\KB$ shows that 59.16\% of the cities are categorized as high popularity, with 29.66\% and 11.18\% categorized as medium and low popularity respectively. To ensure that this skew in the data does not reflect in our generated queries, we use a popularity-based stratification while generating the configurations such that there is an equal representation of low, medium, and popular cities. 

Furthermore, looking into the seasonality reveals that the peak seasons across cities are the summer months in Europe, from May -- September. December also sees a slight increase in seasonality, potentially due to the Christmas holiday season, which precedes the decrease in seasonality over the winter. January -- March is the low-season months across cities. 

Following the classification of the interests into the 5 categories as described in \autoref{subsection: formal description}, we observe that the distribution of travel interests in our \textit{KB} is predominantly related to \textit{Arts \& Entertainment} and \textit{Food}, comprising of almost 70\% of the dataset, followed by \textit{Outdoors \& Recreation}, \textit{Nightlife Spot} and \textit{Shops \& Services}. These interest categories cover a wide range of points of interest, from museums, galleries, and theaters to restaurants, bars, and natural sights, ensuring diversity of travel interests during query generation.

\subsection{Generated Queries} \label{subsection: GenQueries}
Our framework results in a total of 4,604 diverse queries generated from 2,302 key functions in 3 experimental settings ($\queryV$, $\queryPZero$, and $\queryPOne$) for each model namely \gemini and \llama. The dataset consists of the mapping between key functions, retrieved-context, and generated queries. The configuration setting includes the selected persona, city popularity, complexity, and travel filters. 
The retrieved context includes the information used to ground the query ($\filter \in \filters$) and the set of cities from our knowledge base that align with the travel filters ($\context \subset \mathcal{D}$). Additionally, the dataset contains three types of queries, corresponding to each experimental setting.

\subsubsection*{Basic Statistics}
Each query in our generated queries dataset is mapped to an un-ranked list of ground truth cities. To ensure balanced exposure for each city, we stratify our key functions based on the city category described in \autoref{subsection: KB}. This results in a well-proportioned distribution of generated queries: 739 low-popularity cities (32\%), 784 medium-popularity cities (34\%), and 779 high-popularity cities (33\%). In this dataset, we have, on average, 11 cities from our $\KB$ mapped to a generated query.

%% file: sections/eval/0_main.tex
\section{Validation of the Generated Queries} \label{section: validation}

In this section, we present the validation of the generated queries, focusing primarily on assessing their quality. Additionally, we demonstrate their practical utility by leveraging a recommender LLM, denoted as $\RecGenLLM$, to generate travel city recommendations. 

\begin{figure} 
    \includegraphics[width=\columnwidth]{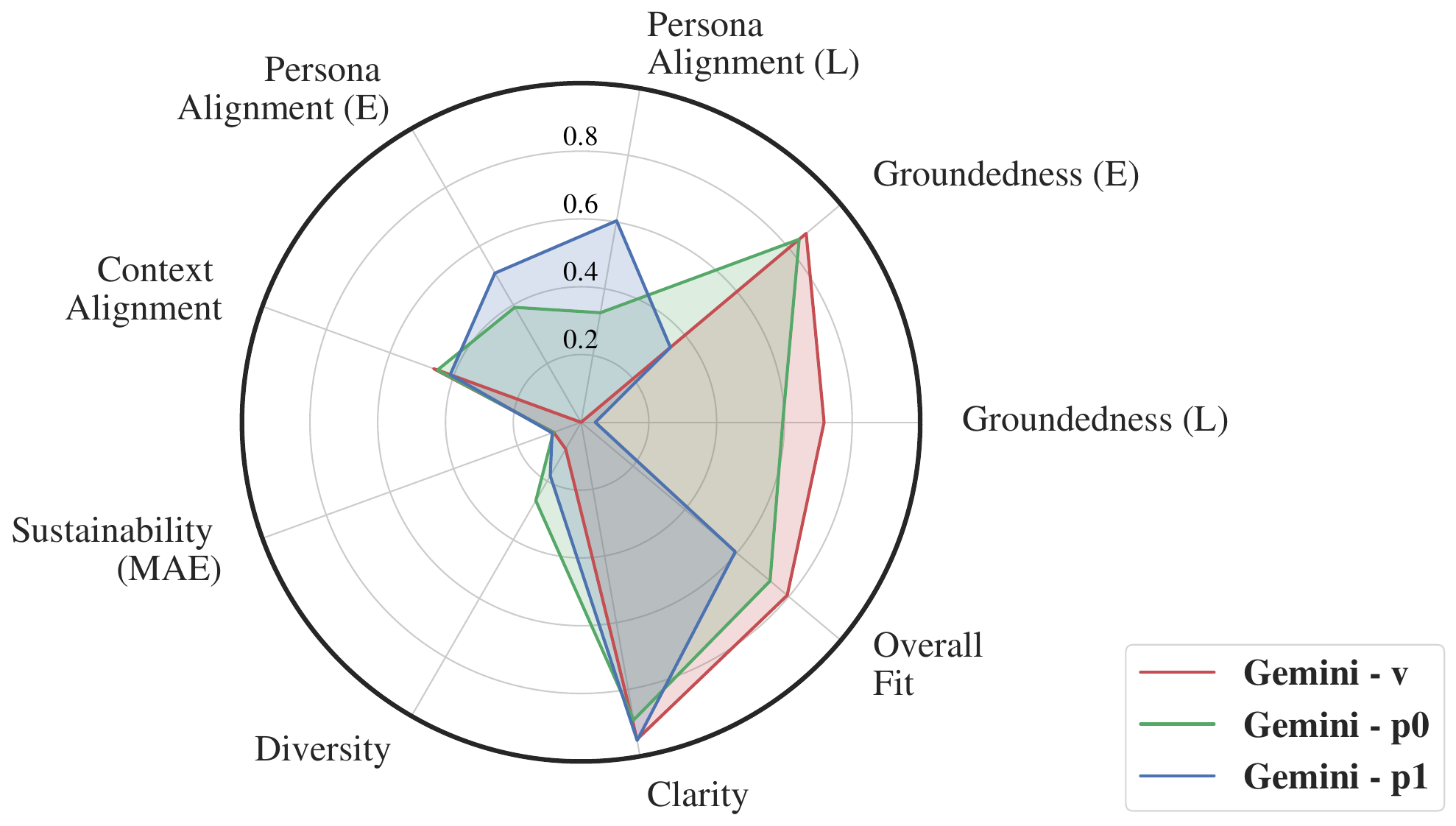}
    \caption{Radar Chart showing the different dimensions of validation and performance of queries generated by \textit{Gemini}. \textit{Llama} shows similar performance across the dimensions and hence is excluded from the paper. L (E) denotes LLM (Expert) validations.}
    \label{fig: radarchart}
\end{figure}

We validate the generated queries across the following dimensions: groundedness with travel filters (\autoref{section: eval_groundedness}), alignment with context (\autoref{section: eval_factcheck}), alignment with personas (\autoref{section: eval_persona_alignment}), adherence to sustainability filters (\autoref{section: eval_sustainability}), diversity of generated queries (\autoref{section: diversity}), and overall clarity and quality assessment (\autoref{section: clarity}). Additionally, \autoref{section: setup} provides an overview of our experimental setup.
~\autoref{fig: radarchart} shows a radar chart showing the different dimensions of validation and performance of queries generated by \textit{Gemini}.

\input{sections/eval/1_setup.tex}

\input{sections/eval/2_filter_groundedness.tex}

\input{sections/eval/4_persona_alignment.tex}
\input{sections/eval/3_factual_groundedness.tex}

\input{sections/eval/5_sustainability.tex}
\input{sections/eval/6_diversity.tex}
\input{sections/eval/7_clarity.tex}

%% file: sections/eval/1_setup.tex
\subsection{Experimental Setup} \label{section: setup}
We use a combination of offline experiments and expert (human) evaluations to assess the quality of the generated queries. 
The subsequent discussion provides a detailed overview of our experimental setup across these validation criteria.
\subsubsection{Offline Experiments}
We assess the quality of the generated queries through the following offline experiments: (1) JudgeLLM, or J-LLM (\gpt) to evaluate alignment with travel filters and persona, (2) semantic similarity for contextual alignment, (3) Mean Average Error (MAE) to measure sustainability alignment, and (4) Self-BLEU scores to assess query diversity.

\subsubsection{Expert Evaluation}

\begin{figure}
    \centering
    \includegraphics[width=\columnwidth]{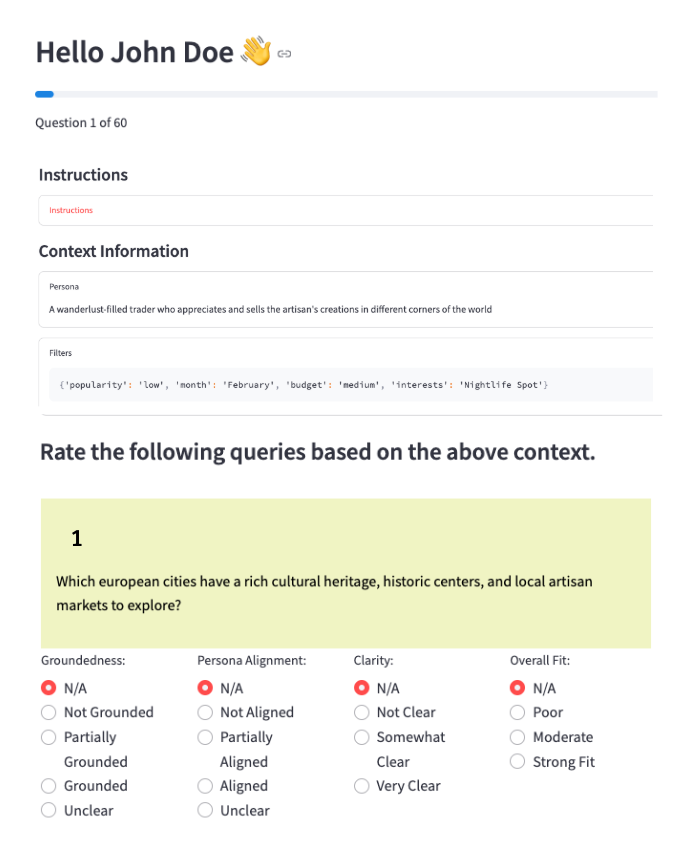}
    \caption{Screenshot showing a part of the evaluation tool developed for the expert study. The full version can be found \href{https://huggingface.co/spaces/ashmib/user-feedback}{\link{here}}.}
    \label{fig: expert eval tool}
\end{figure}

To ensure that the responses generated by the J-LLM align with human judgments, we conduct an expert evaluation using 60 example queries from each model. Two domain experts independently assess these queries using a custom evaluation tool developed specifically for this study. The tool, built with Python's Streamlit library\footnote{https://streamlit.io}, is fully open-source, responsive, and publicly available on Hugging Face Spaces\footnote{https://huggingface.co/spaces} and can be accessed \href{https://huggingface.co/spaces/ashmib/user-feedback}{\link{here}}.
~\autoref{fig: expert eval tool} shows a screenshot from a part of this tool.

The queries generated by all three methods --- $\queryV$, $\queryPZero$, and $\queryPOne$ --- are reviewed by experts using the tool, without knowledge of which model or settings were used for the generation. This approach helps minimize bias. Each expert is also provided with the relevant travel filters and persona for context.
One alternative we considered was to ask users to generate realistic queries themselves based on the filters and travel constraints rather than evaluate system-generated ones. However, this would have required recruiting linguists and domain experts capable of producing high-quality, representative queries --- resources that are both scarce and difficult to source.
Hence, we proceeded with our expert-based evaluation approach. 
Future work could explore this approach to create more naturalistic evaluation benchmarks.

Our interface features detailed instructions on the evaluation process, initially collapsed for a streamlined experience but expandable for full details. Experts are able to submit responses incrementally, save progress, and resume later, with all responses securely stored in a \textbf{Firebase real-time database}~\cite{FirebaseDB}, ensuring seamless retrieval of the evaluation state.
The evaluation focused on multiple dimensions, including alignment with personas, travel filters, clarity, and overall fit, as these aspects are often nuanced and challenging for LLMs to capture.

The following sections provide a detailed analysis of the evaluation outcomes for each query dimension.

%% file: sections/eval/2_filter_groundedness.tex
\subsection{Groundedness with Travel Filters} \label{section: eval_groundedness}

To measure how well the generated queries represent the input constraints or the travel filters, we measure the groundedness using LLMs as an evaluator~\cite{zheng2023judging}. We use \gptfour as our JudgeLLM (J-LLM) to evaluate the queries generated by \llama and \gemini by providing the generated query and the travel filters and asking it to find the number of filters present in the query, along with an explanation. We then use these matches to compute the Mean Recall (MR) for each query setting as follows: 

\begin{equation}
    MR = \sum_{i=1}^{N} \frac{len(matches)_i}{len(filters)_i} 
\end{equation}

The results of the MR scores can be found in~\autoref{tab: llm-eval}. We can see that the vanilla setting ($\queryV$) consistently \update{appears to outperform} the other two settings across models, with the personalized zero-shot ($\queryPZero$) setting showing the second-best performance. However, the personalized single-shot ($\queryPOne$) setting shows extremely low recall values across models. We theorize that this is due to the models overfitting on the persona and the ICL example provided in the prompts, which also led to lower diversity scores for this setting. 

Expert evaluations align with the observed trends of J-LLM in~\autoref{tab: llm-eval}, with evaluators categorizing queries into four levels: 0 for \emph{Unclear}, 1 for \emph{Not grounded}, 2 for \emph{Partially grounded} and 3 for \emph{Grounded}, based on their alignment with travel filters. The percentage of queries rated as \emph{Grounded} (level 3) is reported in~\autoref{tab: llm-eval}.
To further analyze the reliability of expert judgments, we measured inter-evaluator agreement, where we obtained a Mean Absolute Error (MAE) of 0.083, indicating a high level of consistency between the two evaluators.

\begin{table}[ht]
    \caption{Validation results assessing the alignment of generated queries with travel filters ($\filters$) and personas ($\persona$), as evaluated by both J-LLM and domain experts. The values in \textbf{bold} (\textit{italics}) denote the maximum (minimum) values for each model across all the settings --- $\queryV$, $\queryPZero$, and $\queryPOne$.~\maxSign (\minSign) means higher (lower) is better.}
    \label{tab: llm-eval}
    \centering
    \begin{tabular}{ccccccc}
        \toprule
        \multirow{3}{*}{\textbf{Model}}  & 
        \multirow{3}{*}{\textbf{\begin{tabular}[c]{@{}c@{}}Query\\ Setting\end{tabular}}} & 
        \multicolumn{4}{c}{\textbf{Alignment (\%)}~\maxSign} \\ \cmidrule{3-6}
        & & \multicolumn{2}{c}{\textbf{$\filters$}} & \multicolumn{2}{c}{\textbf{$\persona$}} \\ \cmidrule{3-4} \cmidrule(lr){5-6}
        & & J-LLM & Expert & J-LLM & Expert \\ \midrule
        \multirow{3}{*}{\textit{Llama}} & $q_{\text{v}}$  & \maxHighlight{0.759} & \maxHighlight{0.869} & N/A & N/A \\
        & $q_{\text{p0}}$ & 0.644 & 0.806 & \minHighlight{34.348} & \minHighlight{33.333} \\
        & $q_{\text{p1}}$ & \minHighlight{0.043} & \minHighlight{0.346} & \maxHighlight{61.538} & \maxHighlight{59.167} \\ \midrule
        \multirow{3}{*}{\textit{Gemini}} & $q_{\text{v}}$  & \maxHighlight{0.717} & \maxHighlight{0.867} & N/A & N/A \\
        & $q_{\text{p0}}$ & 0.595 & 0.840 & \minHighlight{32.841} & \minHighlight{39.167} \\
        & $q_{\text{p1}}$ & \minHighlight{0.042} & \minHighlight{0.344} & \maxHighlight{60.382} & \maxHighlight{50.833} \\ \bottomrule
    \end{tabular}
\end{table}

%% file: sections/eval/4_persona_alignment.tex
\subsection{Persona Alignment} \label{section: eval_persona_alignment}

We assess the alignment of generated queries with personas using expert evaluation and J-LLM (\gpt), following a two-fold approach similar to that described in~\autoref{section: eval_groundedness}. Both LLM-based and expert evaluators are provided with the persona description, the query, and four rating options: \textit{Not Aligned}, \textit{Partially Aligned}, \textit{Aligned}, and \textit{Unclear}. They are asked to determine how likely the persona would formulate the given query, focusing on tone and phrasing rather than inferred filters.

To mitigate potential biases associated with specific personas, the evaluation prompts explicitly instruct the LLMs and expert evaluators to assess only the tone and linguistic style of the query rather than making assumptions about the persona’s domain expertise or preferences. This ensures that alignment ratings reflect the stylistic coherence between the query and persona rather than pre-existing biases.

\autoref{tab: llm-eval} presents the results of the persona alignment for each model and query setting, comparing J-LLM and expert evaluations. We measure the persona alignment as the percentage of queries rated as \textit{Aligned}. Although $\queryPOne$ shows weaker groundedness with the travel filters, it achieves a higher persona alignment score than $\queryPZero$, according to J-LLM and expert evaluations. This suggests a potential trade-off between persona alignment and query groundedness.

To quantify inter-evaluator agreement, we calculate the MAE score between expert ratings, obtaining a moderate agreement level with an MAE of 0.3. The relatively high MAE suggests that persona interpretation is inherently subjective and can vary across individuals, highlighting the challenges of achieving consistent persona alignment in query generation.

%% file: sections/eval/3_factual_groundedness.tex
\subsection{Contexual Alignment} \label{section: eval_factcheck} 

\begin{table*}[]
    \centering
    \caption{Evaluation of context groundedness, sustainability features, and diversity in the generated queries. The values in \textbf{bold} (\textit{italics}) denote the maximum (minimum) values for each model across all the settings -- $\queryV$, $\queryPZero$, and $\queryPOne$.~\maxSign (\minSign) means higher (lower) is better.}
    \label{tab: offline-evaluations}
    \begin{tabular}{cccccccccc}
    \toprule
    \multirow{3}{*}{\textbf{Model}} & \multirow{3}{*}{\textbf{\begin{tabular}[c]{@{}c@{}}Query\\ Setting\end{tabular}}} & \multirow{3}{*}{\textbf{\begin{tabular}[c]{@{}c@{}}Contextual\\ Alignment \maxSign\end{tabular}}} & \multicolumn{2}{c}{\textbf{Sustainability}} & \multicolumn{5}{c}{\textbf{Diversity \maxSign}} \\ \cmidrule{4-5} \cmidrule(lr){6-10} 
 &  &  & \multirow{2}{*}{\textbf{Similarity \maxSign}} & \multirow{2}{*}{\textbf{MAE \minSign}} & \multirow{2}{*}{\textbf{Overall}} & \multicolumn{4}{c}{\textbf{Query Level}} \\ \cmidrule{7-10} 
 &  &  &  &  &  & \textbf{$\filter_{\text{easy}}$} & \textbf{$\filter_{\text{med}}$} & \textbf{$\filter_{\text{hard}}$} & \textbf{$\filter_{\text{sust}}$} \\ \midrule
    N/A & \textbf{Baseline} & \textcolor{blue}{0.073} & N/A & N/A & - & - & - & - & - \\ \midrule
    \multirow{3}{*}{\textit{Llama}} & $\queryV$ & \maxHighlight{0.482} & \maxHighlight{0.659} & \minHighlight{0.077} & \minHighlight{0.101} & \minHighlight{0.138} & 0.136 & \minHighlight{0.169} & \minHighlight{0.179} \\
     & $\queryPZero$ & 0.455 & 0.624 & 0.078 & \maxHighlight{0.221} & \maxHighlight{0.352} & \maxHighlight{0.274} & 0.299 & 0.319 \\
     & $\queryPOne$ & \minHighlight{0.430} & \minHighlight{0.605} & \maxHighlight{0.091} & 0.177 & 0.307 & \minHighlight{0.241} & \maxHighlight{0.315} & \maxHighlight{0.327} \\ \midrule
    \multirow{3}{*}{\textit{Gemini}} & $\queryV$ & \maxHighlight{0.463} & \maxHighlight{0.648} & \minHighlight{0.086} & \minHighlight{0.091} & \minHighlight{0.140} & \minHighlight{0.119} & \minHighlight{0.113} & \minHighlight{0.167} \\
     & $\queryPZero$ & 0.451 & 0.589 & 0.087 & \maxHighlight{0.267} & \maxHighlight{0.415} & \maxHighlight{0.321} & 0.327 & 0.360 \\
     & $\queryPOne$ & \minHighlight{0.411} & \minHighlight{0.526} & \maxHighlight{0.091} & 0.182 & 0.346 & 0.251 & \maxHighlight{0.342} & \maxHighlight{0.378} \\ \bottomrule
    \end{tabular}
    \end{table*}

Besides evaluating alignment with travel filters and personas, we assess query groundedness within the context to minimize LLM hallucinations and ensure factual accuracy from our $\KB$.
To quantify this, we perform a semantic search for the query \textit{q} using our $\KB$ (now represented as a vector database) to retrieve the query context and cities, respectively. Our assumption follows that if the queries are factually grounded in the $\KB$, then a subsequent semantic retrieval using these queries from the same $\KB$ should result in the same information. This retrieved context is compared with the original context used during query generation by computing the cosine similarity between the respective contextual embeddings generated using the \textit{all-mpnet-base-v2}\footnote{\label{mpnet}https://huggingface.co/sentence-transformers/all-mpnet-base-v2} model. 

From \autoref{tab: offline-evaluations}, we observe that the models and query settings show similar performance across metrics, with the vanilla setting \update{appearing to barely outperform}the other two settings. However, there is a sizeable increase in performance when compared to the baseline, which is computed by comparing a paragraph of Lorem Ipsum text with the original context, indicating that the addition of the $\KB$ in the query generation leads to queries that are factually grounded.

%% file: sections/eval/5_sustainability.tex
\subsection{Sustainability Alignment} \label{section: eval_sustainability}

The \textbf{\textit{sustainable}} query complexity level $\filter_\text{sust}$ 
contains two non-sustainable travel filters and one sustainable filter, along with the destination popularity information. 
While we want to introduce sustainable travel into our dataset in a seamless manner, it is also important to verify that the sustainable queries do not show substantial differences semantically when compared to the other queries. In order to measure this, we compute the cosine similarity scores using the \textit{all-mpnet-base-v2} model between the sustainable queries and non-sustainable queries, as shown in~\autoref{tab: offline-evaluations}. 

The results show moderately high scores between the sustainable and non-sustainable queries across model and query settings, indicating semantic similarity between the two sets. However, this does not indicate whether the sustainable filter is well represented within each sustainable query or if it is compromised in favor of the non-sustainable travel filters. 

To evaluate how well the generated queries consider the sustainability constraints, we compute the cosine similarities between each travel filter and the query belonging to the sustainable complexity ($q_i$). We further compute the MAE between the similarities of the sustainability and the average similarity of the non-sustainability filters (represented by \textit{j}) for each query setting as the following: 

\begin{equation}
    \text{MAE} = \frac{1}{N} \sum_{i=1}^{N} \left| sim_{\text{sust}, q_i} - \frac{1}{M} \sum_{j=1}^{M} sim_{j, q_i} \right|
\end{equation}

The results in~\autoref{tab: offline-evaluations} show low MAE scores across query settings and models, suggesting that the LLMs do not overlook sustainability in favor of the other travel filters, ensuring a balanced representation of sustainability during query generation.

%% file: sections/eval/6_diversity.tex
\subsection{Diversity of the Generated Queries} \label{section: diversity}
Given that the filter and context control the sentence semantics, we measure the diversity of the generated queries at the token level. We use Self-BLEU~\cite{zhu2018texygen}, a metric derived from the BLEU score~\cite{papineni2002bleu}, which calculates sentence similarity between a hypothesis and a set of reference texts using n-gram matching. For each generated query $q_\text{i}$, we measure Self-BLEU against the other queries as the reference set using 4-grams. We define the diversity score as 
\begin{equation}
    Diversity = 1 - \frac{(\Sigma^{|G|}_{i=1} Self\text{-}BLEU(q_i))}{|G|}, \quad q \in G, G \subseteq \Query
\end{equation}
A higher diversity score signals better diversity.

~\autoref{tab: offline-evaluations} shows that personalization improves query diversity, with $q_{\text{p0}}$ and $q_{\text{p1}}$ yield higher scores than $q_{\text{v}}$. We observe that queries generated with the same persona tend to sound similar across different key functions. This resulted in a lower overall diversity score since each persona is assigned to 12 key functions. When analyzing the scores at the query level, it becomes apparent --- especially for sustainable queries, adding more contextual dimensions leads to higher diversity. However, potential overfitting in $q_{\text{p1}}$ suggests that queries become less diverse when closely mirroring examples in the prompt. This underscores the need for careful prompt design to balance personalization and diversity in generated queries.

%% file: sections/eval/7_clarity.tex
\subsection{Clarity and Overall Quality Assessment} \label{section: clarity}
While the J-LLM evaluations can assess most aspects of query quality, they often fail to capture subtle nuances in language and semantics. For instance, consider the following persona:  
\emph{"A franchise owner of a major retail brand planning to establish a branch in the town,"}
with a preference for low popularity and low budget. The generated query for this persona, $q_{\text{p1}}$, is: \emph{"Low-cost European city with untapped market potential for retail."}

An LLM-based evaluation assigns this query a perfect persona alignment and groundedness score. However, expert evaluators highlight an important nuance: the phrase \emph{"untapped market potential for retail"} is not directly aligned with the stated preferences. Consequently, they assign a moderate overall fit score rather than a perfect one. This discrepancy underscores the importance of expert evaluations, particularly in assessing clarity and overall fit.  
Hence, we measure two additional aspects—\emph{clarity} and \emph{overall fit}—through expert review. \emph{Clarity} refers to how grammatically correct, understandable, and interpretable a query is, while \emph{overall fit} captures a holistic assessment considering relevance, clarity, and persona alignment. 

For \emph{clarity}, the average score, normalized over the scale of evaluation choices, is high across models and settings, with $\queryPOne$ showing the highest scores at 0.947 and 0.953 for \llama and \gemini, respectively. Furthermore, the mean absolute error (MAE) for clarity ratings across two independent evaluators is extremely low (0.011), indicating a high degree of agreement.  

Similarly, evaluators provide consistent ratings across all queries and models for overall fit, averaging 0.7 with minimal deviation. However, we see the opposite trends in \emph{overall fit} as compared to~\emph{clarity}, with $\queryV$ having the highest scores in both models, likely due to the overfitting of the $\queryPOne$ setting to the persona, indicating a correlation between the \emph{overall fit} and other validation criteria, such as groundedness in filters and context. The intra-evaluator MAE for overall fit is 0.16, further confirming the reliability of their judgments. These results reinforce the necessity of expert evaluation to capture subtleties overlooked by automated methods.

%% file: sections/recllm_eval.tex
\subsection{Preliminary Analysis using RecGenLLM}

As a preliminary analysis, we evaluated the queries generated by Gemini as $\QueryGenLLM$ using the same Gemini model (\gemini), now serving as our recommender LLM (RecGenLLM, $\RecGenLLM$), grounded with web search~\cite{websearch}. 
While query generation and recommendation were performed using \gemini in our setup, other LLMs could also function as $\RecGenLLM$, as their generative capabilities are inherently task- and prompt-dependent~\cite{machlab2024llm}. 
We instructed RecGenLLM to recommend cities from the predefined 200 cities in our $\KB$, in zero- and few-shot settings. Grounding it with web search allowed it to provide citations from web searches if available.

On average, RecGenLLM recommends 10 cities per query, of which eight cities (80\%) are found in our knowledge base (KB). 
To assess the impact of popularity bias, we computed the percentage of recommended cities classified as "high-popularity" based on the popularity scores in our $\KB$ and averaged this across all queries. On average, 79\% of the cities recommended by $\RecGenLLM$ are highly popular, regardless of the query constraints. However, on a closer look, we observe that the $\RecGenLLM$ tends to recommend popular cities 77\% of the time, even when the queries specifically request for cities with low popularity. For example, in July, the $\recGenCities$ for the query \textit{``Trip to a less-visited European city with historical sites and museums. High budget.''} include cities like \textit{Budapest (Hungary) and Prague (Czechia)}, which are not only popular destinations but also have peak season in July. In contrast, the $c_\context$ for this query consists of less popular cities like \textit{Syktyvkar (Russia), Malatya and Kars (Türkiye), Ioannina (Greece)}.

This behavior is likely due to the limited capability of the $\RecGenLLM$ to understand nuanced queries and predilection to recommend more common travel destinations. Addressing this issue in future work will require developing retrieval strategies for RecGenLLM that actively mitigate popularity bias, ensuring more balanced and diverse recommendations~\cite{lichtenberg2024large, guo2024bias}.

%% file: sections/conclusion.tex
\section{Conclusion} \label{section: conclusion}

In conclusion, this paper introduces SynthTRIPs, a novel framework leveraging LLMs and a curated knowledge base for generating synthetic travel queries, addressing the need for more personalized and sustainable travel recommendation datasets. 
Through persona-based preferences, sustainability filters, and grounding techniques, SynthTRIPs effectively creates realistic and diverse queries, as validated by human experts and automatic LLM-based assessments. 
The framework's reproducible pipeline, open-source resources, and flexible design offer valuable tools for researchers in the tourism recommendation community. 

However, SynthTRIPs has limitations too, including the potential for subjective persona interpretation and the observed trade-off between personalization and groundedness, where highly personalized queries may exhibit lower factual accuracy or alignment with their constraints. 
Furthermore, the personas in SynthTRIPs are not always domain-aligned. Since PersonaHub consists of synthetically generated data, it often lacks diversity and realism. While clustering them by topic marginally improves diversity, the issue of unrealism in the personas persists and remains evident in our dataset.

Furthermore, our preliminary analysis reveals a potential popularity bias in a zero/few-shot setting through a RecGenLLM to generate city recommendations. In this case, the model often favors high-popularity destinations, even when users explicitly specify a preference for less popular options. Since SynthTRIPs ensures an equal representation of popular and less popular travel destinations, future work could use this to focus on mitigating popularity bias in zero/few-shot recommendations and developing more sophisticated retrieval strategies to ensure balanced and diverse results.
Future work could extend SynthTRIPs by incorporating more user filters (e.g., traveling with families, solo travels, accessibility, etc.), expanding beyond European cities to enhance global applicability, and developing strategies to maintain the knowledge base as real-world travel data evolves.

%% file: main.bbl

\begin{thebibliography}{43}


\ifx \showCODEN    \undefined \def \showCODEN     #1{\unskip}     \fi
\ifx \showISBNx    \undefined \def \showISBNx     #1{\unskip}     \fi
\ifx \showISBNxiii \undefined \def \showISBNxiii  #1{\unskip}     \fi
\ifx \showISSN     \undefined \def \showISSN      #1{\unskip}     \fi
\ifx \showLCCN     \undefined \def \showLCCN      #1{\unskip}     \fi
\ifx \shownote     \undefined \def \shownote      #1{#1}          \fi
\ifx \showarticletitle \undefined \def \showarticletitle #1{#1}   \fi
\ifx \showURL      \undefined \def \showURL       {\relax}        \fi
\providecommand\bibfield[2]{#2}
\providecommand\bibinfo[2]{#2}
\providecommand\natexlab[1]{#1}
\providecommand\showeprint[2][]{arXiv:#2}

\bibitem[Abbasiantaeb et~al\mbox{.}(2024)]%
        {abbasiantaeb2024let}
\bibfield{author}{\bibinfo{person}{Zahra Abbasiantaeb}, \bibinfo{person}{Yifei
  Yuan}, \bibinfo{person}{Evangelos Kanoulas}, {and} \bibinfo{person}{Mohammad
  Aliannejadi}.} \bibinfo{year}{2024}\natexlab{}.
\newblock \showarticletitle{Let the llms talk: Simulating human-to-human
  conversational qa via zero-shot llm-to-llm interactions}. In
  \bibinfo{booktitle}{\emph{Proceedings of the 17th ACM International
  Conference on Web Search and Data Mining}}. \bibinfo{pages}{8--17}.
\newblock


\bibitem[Ashby et~al\mbox{.}(2024)]%
        {ashby2024towards}
\bibfield{author}{\bibinfo{person}{Trevor Ashby}, \bibinfo{person}{Adithya
  Kulkarni}, \bibinfo{person}{Jingyuan Qi}, \bibinfo{person}{Minqian Liu},
  \bibinfo{person}{Eunah Cho}, \bibinfo{person}{Vaibhav Kumar}, {and}
  \bibinfo{person}{Lifu Huang}.} \bibinfo{year}{2024}\natexlab{}.
\newblock \showarticletitle{Towards Effective Long Conversation Generation with
  Dynamic Topic Tracking and Recommendation}. In
  \bibinfo{booktitle}{\emph{Proceedings of the 17th International Natural
  Language Generation Conference}}. \bibinfo{pages}{540--556}.
\newblock


\bibitem[Balakrishnan and W{\"o}rndl(2021)]%
        {balakrishnan2021multistakeholder}
\bibfield{author}{\bibinfo{person}{Gokulakrishnan Balakrishnan} {and}
  \bibinfo{person}{Wolfgang W{\"o}rndl}.} \bibinfo{year}{2021}\natexlab{}.
\newblock \showarticletitle{Multistakeholder Recommender Systems in Tourism}.
\newblock \bibinfo{journal}{\emph{Proc. Workshop on Recommenders in Tourism
  (RecTour 2021)}} (\bibinfo{year}{2021}).
\newblock


\bibitem[Banerjee et~al\mbox{.}(2025)]%
        {banerjee2025modeling}
\bibfield{author}{\bibinfo{person}{Ashmi Banerjee}, \bibinfo{person}{Tunar
  Mahmudov}, \bibinfo{person}{Emil Adler}, \bibinfo{person}{Fitri~Nur Aisyah},
  {and} \bibinfo{person}{Wolfgang W{\"o}rndl}.}
  \bibinfo{year}{2025}\natexlab{}.
\newblock \showarticletitle{Modeling sustainable city trips: integrating CO 2 e
  emissions, popularity, and seasonality into tourism recommender systems}.
\newblock \bibinfo{journal}{\emph{Information Technology \& Tourism}}
  (\bibinfo{year}{2025}), \bibinfo{pages}{1--38}.
\newblock


\bibitem[Banerjee et~al\mbox{.}(2024)]%
        {banerjee2024enhancing}
\bibfield{author}{\bibinfo{person}{Ashmi Banerjee}, \bibinfo{person}{Adithi
  Satish}, {and} \bibinfo{person}{Wolfgang W{\"o}rndl}.}
  \bibinfo{year}{2024}\natexlab{}.
\newblock \showarticletitle{Enhancing Tourism Recommender Systems for
  Sustainable City Trips Using Retrieval-Augmented Generation}.
\newblock \bibinfo{journal}{\emph{arXiv preprint arXiv:2409.18003}}
  (\bibinfo{year}{2024}).
\newblock


\bibitem[Chaudhari and Thakkar(2020)]%
        {chaudhari2020comprehensive}
\bibfield{author}{\bibinfo{person}{Kinjal Chaudhari} {and}
  \bibinfo{person}{Ankit Thakkar}.} \bibinfo{year}{2020}\natexlab{}.
\newblock \showarticletitle{A comprehensive survey on travel recommender
  systems}.
\newblock \bibinfo{journal}{\emph{Archives of computational methods in
  engineering}}  \bibinfo{volume}{27} (\bibinfo{year}{2020}),
  \bibinfo{pages}{1545--1571}.
\newblock


\bibitem[Deldjoo(2023)]%
        {deldjoo2023fairness}
\bibfield{author}{\bibinfo{person}{Yashar Deldjoo}.}
  \bibinfo{year}{2023}\natexlab{}.
\newblock \showarticletitle{Fairness of chatgpt and the role of
  explainable-guided prompts}. In \bibinfo{booktitle}{\emph{Joint European
  Conference on Machine Learning and Knowledge Discovery in Databases}}.
  Springer, \bibinfo{pages}{13--22}.
\newblock


\bibitem[Firebase(2025)]%
        {FirebaseDB}
\bibfield{author}{\bibinfo{person}{Google Firebase}.}
  \bibinfo{year}{2025}\natexlab{}.
\newblock \bibinfo{booktitle}{\emph{Firebase Realtime Database Documentation}}.
\newblock
\urldef\tempurl%
\url{https://firebase.google.com/docs/database}
\showURL{%
\tempurl}


\bibitem[Fr{\"o}hling et~al\mbox{.}(2024)]%
        {frohling2024personas}
\bibfield{author}{\bibinfo{person}{Leon Fr{\"o}hling},
  \bibinfo{person}{Gianluca Demartini}, {and} \bibinfo{person}{Dennis
  Assenmacher}.} \bibinfo{year}{2024}\natexlab{}.
\newblock \showarticletitle{Personas with Attitudes: Controlling LLMs for
  Diverse Data Annotation}.
\newblock \bibinfo{journal}{\emph{arXiv preprint arXiv:2410.11745}}
  (\bibinfo{year}{2024}).
\newblock


\bibitem[Ge et~al\mbox{.}(2024)]%
        {ge2024scaling}
\bibfield{author}{\bibinfo{person}{Tao Ge}, \bibinfo{person}{Xin Chan},
  \bibinfo{person}{Xiaoyang Wang}, \bibinfo{person}{Dian Yu},
  \bibinfo{person}{Haitao Mi}, {and} \bibinfo{person}{Dong Yu}.}
  \bibinfo{year}{2024}\natexlab{}.
\newblock \showarticletitle{Scaling synthetic data creation with 1,000,000,000
  personas}.
\newblock \bibinfo{journal}{\emph{arXiv preprint arXiv:2406.20094}}
  (\bibinfo{year}{2024}).
\newblock


\bibitem[Google({[n.\,d.]})]%
        {websearch}
\bibfield{author}{\bibinfo{person}{Google}.}
  \bibinfo{year}{[n.\,d.]}\natexlab{}.
\newblock \bibinfo{booktitle}{\emph{Ground Gemini to your data}}.
\newblock
\urldef\tempurl%
\url{https://cloud.google.com/vertex-ai/generative-ai/docs/multimodal/ground-with-your-data#private-ground-gemini}
\showURL{%
\tempurl}


\bibitem[Grootendorst(2022)]%
        {grootendorst2022bertopic}
\bibfield{author}{\bibinfo{person}{Maarten Grootendorst}.}
  \bibinfo{year}{2022}\natexlab{}.
\newblock \showarticletitle{BERTopic: Neural topic modeling with a class-based
  TF-IDF procedure}.
\newblock \bibinfo{journal}{\emph{arXiv preprint arXiv:2203.05794}}
  (\bibinfo{year}{2022}).
\newblock


\bibitem[Guo et~al\mbox{.}(2024)]%
        {guo2024bias}
\bibfield{author}{\bibinfo{person}{Yufei Guo}, \bibinfo{person}{Muzhe Guo},
  \bibinfo{person}{Juntao Su}, \bibinfo{person}{Zhou Yang},
  \bibinfo{person}{Mengqiu Zhu}, \bibinfo{person}{Hongfei Li},
  \bibinfo{person}{Mengyang Qiu}, {and} \bibinfo{person}{Shuo~Shuo Liu}.}
  \bibinfo{year}{2024}\natexlab{}.
\newblock \showarticletitle{Bias in large language models: Origin, evaluation,
  and mitigation}.
\newblock \bibinfo{journal}{\emph{arXiv preprint arXiv:2411.10915}}
  (\bibinfo{year}{2024}).
\newblock


\bibitem[Haussmann et~al\mbox{.}(2019)]%
        {haussmann2019foodkg}
\bibfield{author}{\bibinfo{person}{Steven Haussmann}, \bibinfo{person}{Oshani
  Seneviratne}, \bibinfo{person}{Yu Chen}, \bibinfo{person}{Yarden Ne’eman},
  \bibinfo{person}{James Codella}, \bibinfo{person}{Ching-Hua Chen},
  \bibinfo{person}{Deborah~L McGuinness}, {and} \bibinfo{person}{Mohammed~J
  Zaki}.} \bibinfo{year}{2019}\natexlab{}.
\newblock \showarticletitle{FoodKG: a semantics-driven knowledge graph for food
  recommendation}. In \bibinfo{booktitle}{\emph{The Semantic Web--ISWC 2019:
  18th International Semantic Web Conference, Auckland, New Zealand, October
  26--30, 2019, Proceedings, Part II 18}}. Springer, \bibinfo{pages}{146--162}.
\newblock


\bibitem[Huang et~al\mbox{.}(2024)]%
        {huang2024survey}
\bibfield{author}{\bibinfo{person}{Lei Huang}, \bibinfo{person}{Weijiang Yu},
  \bibinfo{person}{Weitao Ma}, \bibinfo{person}{Weihong Zhong},
  \bibinfo{person}{Zhangyin Feng}, \bibinfo{person}{Haotian Wang},
  \bibinfo{person}{Qianglong Chen}, \bibinfo{person}{Weihua Peng},
  \bibinfo{person}{Xiaocheng Feng}, \bibinfo{person}{Bing Qin},
  {et~al\mbox{.}}} \bibinfo{year}{2024}\natexlab{}.
\newblock \showarticletitle{A survey on hallucination in large language models:
  Principles, taxonomy, challenges, and open questions}.
\newblock \bibinfo{journal}{\emph{ACM Transactions on Information Systems}}
  (\bibinfo{year}{2024}).
\newblock


\bibitem[Isinkaye et~al\mbox{.}(2015)]%
        {isinkaye2015recommendation}
\bibfield{author}{\bibinfo{person}{Folasade~Olubusola Isinkaye},
  \bibinfo{person}{Yetunde~O Folajimi}, {and}
  \bibinfo{person}{Bolande~Adefowoke Ojokoh}.} \bibinfo{year}{2015}\natexlab{}.
\newblock \showarticletitle{Recommendation systems: Principles, methods and
  evaluation}.
\newblock \bibinfo{journal}{\emph{Egyptian informatics journal}}
  \bibinfo{volume}{16}, \bibinfo{number}{3} (\bibinfo{year}{2015}),
  \bibinfo{pages}{261--273}.
\newblock


\bibitem[Jandaghi et~al\mbox{.}(2023)]%
        {jandaghi2023faithful}
\bibfield{author}{\bibinfo{person}{Pegah Jandaghi}, \bibinfo{person}{XiangHai
  Sheng}, \bibinfo{person}{Xinyi Bai}, \bibinfo{person}{Jay Pujara}, {and}
  \bibinfo{person}{Hakim Sidahmed}.} \bibinfo{year}{2023}\natexlab{}.
\newblock \showarticletitle{Faithful persona-based conversational dataset
  generation with large language models}.
\newblock \bibinfo{journal}{\emph{arXiv preprint arXiv:2312.10007}}
  (\bibinfo{year}{2023}).
\newblock


\bibitem[Jannach and Chen(2022)]%
        {jannach2022conversational}
\bibfield{author}{\bibinfo{person}{Dietmar Jannach} {and} \bibinfo{person}{Li
  Chen}.} \bibinfo{year}{2022}\natexlab{}.
\newblock \showarticletitle{Conversational recommendation: A grand AI
  challenge}.
\newblock \bibinfo{journal}{\emph{AI Magazine}} \bibinfo{volume}{43},
  \bibinfo{number}{2} (\bibinfo{year}{2022}), \bibinfo{pages}{151--163}.
\newblock


\bibitem[Khalil et~al\mbox{.}(2025)]%
        {khalil2025creating}
\bibfield{author}{\bibinfo{person}{Mohammad Khalil}, \bibinfo{person}{Farhad
  Vadiee}, \bibinfo{person}{Ronas Shakya}, {and} \bibinfo{person}{Qinyi Liu}.}
  \bibinfo{year}{2025}\natexlab{}.
\newblock \showarticletitle{Creating Artificial Students that Never Existed:
  Leveraging Large Language Models and CTGANs for Synthetic Data Generation}.
\newblock \bibinfo{journal}{\emph{arXiv preprint arXiv:2501.01793}}
  (\bibinfo{year}{2025}).
\newblock


\bibitem[Lan et~al\mbox{.}(2022)]%
        {lan_knowledge_2022}
\bibfield{author}{\bibinfo{person}{Jian Lan}, \bibinfo{person}{Runfeng Shi},
  \bibinfo{person}{Ye Cao}, {and} \bibinfo{person}{Jiancheng Lv}.}
  \bibinfo{year}{2022}\natexlab{}.
\newblock \showarticletitle{Knowledge {Graph}-based {Conversational}
  {Recommender} {System} in {Travel}}. In \bibinfo{booktitle}{\emph{2022
  {International} {Joint} {Conference} on {Neural} {Networks} ({IJCNN})}}.
  \bibinfo{pages}{1--8}.
\newblock
\href{https://doi.org/10.1109/IJCNN55064.2022.9892176}{doi:\nolinkurl{10.1109/IJCNN55064.2022.9892176}}
\newblock
\shownote{ISSN: 2161-4407}.


\bibitem[Leszczynski et~al\mbox{.}(2023)]%
        {leszczynski2023talk}
\bibfield{author}{\bibinfo{person}{Megan Leszczynski}, \bibinfo{person}{Shu
  Zhang}, \bibinfo{person}{Ravi Ganti}, \bibinfo{person}{Krisztian Balog},
  \bibinfo{person}{Filip Radlinski}, \bibinfo{person}{Fernando Pereira}, {and}
  \bibinfo{person}{Arun~Tejasvi Chaganty}.} \bibinfo{year}{2023}\natexlab{}.
\newblock \showarticletitle{Talk the Walk: Synthetic Data Generation for
  Conversational Music Recommendation}.
\newblock \bibinfo{journal}{\emph{arXiv preprint arXiv:2301.11489}}
  (\bibinfo{year}{2023}).
\newblock


\bibitem[Lewis(2019)]%
        {lewis2019bart}
\bibfield{author}{\bibinfo{person}{Mike Lewis}.}
  \bibinfo{year}{2019}\natexlab{}.
\newblock \showarticletitle{Bart: Denoising sequence-to-sequence pre-training
  for natural language generation, translation, and comprehension}.
\newblock \bibinfo{journal}{\emph{arXiv preprint arXiv:1910.13461}}
  (\bibinfo{year}{2019}).
\newblock


\bibitem[Lichtenberg et~al\mbox{.}(2024)]%
        {lichtenberg2024large}
\bibfield{author}{\bibinfo{person}{Jan~Malte Lichtenberg},
  \bibinfo{person}{Alexander Buchholz}, {and} \bibinfo{person}{Pola
  Schw{\"o}bel}.} \bibinfo{year}{2024}\natexlab{}.
\newblock \showarticletitle{Large language models as recommender systems: A
  study of popularity bias}.
\newblock \bibinfo{journal}{\emph{arXiv preprint arXiv:2406.01285}}
  (\bibinfo{year}{2024}).
\newblock


\bibitem[Lim et~al\mbox{.}(2018)]%
        {lim2018personalized}
\bibfield{author}{\bibinfo{person}{Kwan~Hui Lim}, \bibinfo{person}{Jeffrey
  Chan}, \bibinfo{person}{Christopher Leckie}, {and} \bibinfo{person}{Shanika
  Karunasekera}.} \bibinfo{year}{2018}\natexlab{}.
\newblock \showarticletitle{Personalized trip recommendation for tourists based
  on user interests, points of interest visit durations and visit recency}.
\newblock \bibinfo{journal}{\emph{Knowledge and Information Systems}}
  \bibinfo{volume}{54} (\bibinfo{year}{2018}), \bibinfo{pages}{375--406}.
\newblock


\bibitem[Lu et~al\mbox{.}(2012)]%
        {lu2012personalized}
\bibfield{author}{\bibinfo{person}{Eric Hsueh-Chan Lu},
  \bibinfo{person}{Ching-Yu Chen}, {and} \bibinfo{person}{Vincent~S Tseng}.}
  \bibinfo{year}{2012}\natexlab{}.
\newblock \showarticletitle{Personalized trip recommendation with multiple
  constraints by mining user check-in behaviors}. In
  \bibinfo{booktitle}{\emph{Proceedings of the 20th International Conference on
  Advances in Geographic Information Systems}}. \bibinfo{pages}{209--218}.
\newblock


\bibitem[Lu et~al\mbox{.}(2023)]%
        {lu2023august}
\bibfield{author}{\bibinfo{person}{Yu Lu}, \bibinfo{person}{Junwei Bao},
  \bibinfo{person}{Zichen Ma}, \bibinfo{person}{Xiaoguang Han},
  \bibinfo{person}{Youzheng Wu}, \bibinfo{person}{Shuguang Cui}, {and}
  \bibinfo{person}{Xiaodong He}.} \bibinfo{year}{2023}\natexlab{}.
\newblock \showarticletitle{AUGUST: an automatic generation understudy for
  synthesizing conversational recommendation datasets}.
\newblock \bibinfo{journal}{\emph{arXiv preprint arXiv:2306.09631}}
  (\bibinfo{year}{2023}).
\newblock


\bibitem[Machlab and Battle(2024)]%
        {machlab2024llm}
\bibfield{author}{\bibinfo{person}{Daniel Machlab} {and} \bibinfo{person}{Rick
  Battle}.} \bibinfo{year}{2024}\natexlab{}.
\newblock \showarticletitle{LLM In-Context Recall is Prompt Dependent}.
\newblock \bibinfo{journal}{\emph{arXiv preprint arXiv:2404.08865}}
  (\bibinfo{year}{2024}).
\newblock


\bibitem[Mar{\i}n et~al\mbox{.}(2021)]%
        {marin2021recipe1m+}
\bibfield{author}{\bibinfo{person}{Javier Mar{\i}n}, \bibinfo{person}{Aritro
  Biswas}, \bibinfo{person}{Ferda Ofli}, \bibinfo{person}{Nicholas Hynes},
  \bibinfo{person}{Amaia Salvador}, \bibinfo{person}{Yusuf Aytar},
  \bibinfo{person}{Ingmar Weber}, {and} \bibinfo{person}{Antonio Torralba}.}
  \bibinfo{year}{2021}\natexlab{}.
\newblock \showarticletitle{Recipe1m+: A dataset for learning cross-modal
  embeddings for cooking recipes and food images}.
\newblock \bibinfo{journal}{\emph{IEEE Transactions on Pattern Analysis and
  Machine Intelligence}} \bibinfo{volume}{43}, \bibinfo{number}{1}
  (\bibinfo{year}{2021}), \bibinfo{pages}{187--203}.
\newblock


\bibitem[Meyer et~al\mbox{.}(2024)]%
        {meyer2024comparison}
\bibfield{author}{\bibinfo{person}{Sonia Meyer}, \bibinfo{person}{Shreya
  Singh}, \bibinfo{person}{Bertha Tam}, \bibinfo{person}{Christopher Ton},
  {and} \bibinfo{person}{Angel Ren}.} \bibinfo{year}{2024}\natexlab{}.
\newblock \showarticletitle{A Comparison of LLM Finetuning Methods \&
  Evaluation Metrics with Travel Chatbot Use Case}.
\newblock \bibinfo{journal}{\emph{arXiv preprint arXiv:2408.03562}}
  (\bibinfo{year}{2024}).
\newblock


\bibitem[Papineni et~al\mbox{.}(2002)]%
        {papineni2002bleu}
\bibfield{author}{\bibinfo{person}{Kishore Papineni}, \bibinfo{person}{Salim
  Roukos}, \bibinfo{person}{Todd Ward}, {and} \bibinfo{person}{Wei-Jing Zhu}.}
  \bibinfo{year}{2002}\natexlab{}.
\newblock \showarticletitle{Bleu: a method for automatic evaluation of machine
  translation}. In \bibinfo{booktitle}{\emph{Proceedings of the 40th annual
  meeting of the Association for Computational Linguistics}}.
  \bibinfo{pages}{311--318}.
\newblock


\bibitem[Sanchez and Dietz(2022)]%
        {sanchez2022travelers}
\bibfield{author}{\bibinfo{person}{Pablo Sanchez} {and}
  \bibinfo{person}{Linus~W Dietz}.} \bibinfo{year}{2022}\natexlab{}.
\newblock \showarticletitle{Travelers vs. Locals: The Effect of Cluster
  Analysis in Point-of-Interest Recommendation}. In
  \bibinfo{booktitle}{\emph{Proceedings of the 30th ACM Conference on User
  Modeling, Adaptation and Personalization}}. \bibinfo{pages}{132--142}.
\newblock


\bibitem[Tang et~al\mbox{.}(2023)]%
        {tang2023does}
\bibfield{author}{\bibinfo{person}{Ruixiang Tang}, \bibinfo{person}{Xiaotian
  Han}, \bibinfo{person}{Xiaoqian Jiang}, {and} \bibinfo{person}{Xia Hu}.}
  \bibinfo{year}{2023}\natexlab{}.
\newblock \showarticletitle{Does synthetic data generation of llms help
  clinical text mining?}
\newblock \bibinfo{journal}{\emph{arXiv preprint arXiv:2303.04360}}
  (\bibinfo{year}{2023}).
\newblock


\bibitem[Thakur et~al\mbox{.}(2021)]%
        {thakur2021beir}
\bibfield{author}{\bibinfo{person}{Nandan Thakur}, \bibinfo{person}{Nils
  Reimers}, \bibinfo{person}{Andreas R{\"u}ckl{\'e}}, \bibinfo{person}{Abhishek
  Srivastava}, {and} \bibinfo{person}{Iryna Gurevych}.}
  \bibinfo{year}{2021}\natexlab{}.
\newblock \showarticletitle{Beir: A heterogenous benchmark for zero-shot
  evaluation of information retrieval models}.
\newblock \bibinfo{journal}{\emph{arXiv preprint arXiv:2104.08663}}
  (\bibinfo{year}{2021}).
\newblock


\bibitem[Wang et~al\mbox{.}(2024)]%
        {wang2024feb4rag}
\bibfield{author}{\bibinfo{person}{Shuai Wang}, \bibinfo{person}{Ekaterina
  Khramtsova}, \bibinfo{person}{Shengyao Zhuang}, {and} \bibinfo{person}{Guido
  Zuccon}.} \bibinfo{year}{2024}\natexlab{}.
\newblock \showarticletitle{Feb4rag: Evaluating federated search in the context
  of retrieval augmented generation}. In \bibinfo{booktitle}{\emph{Proceedings
  of the 47th International ACM SIGIR Conference on Research and Development in
  Information Retrieval}}. \bibinfo{pages}{763--773}.
\newblock


\bibitem[Wang et~al\mbox{.}(2023)]%
        {wang2023improving}
\bibfield{author}{\bibinfo{person}{Xi Wang}, \bibinfo{person}{Hossein~A
  Rahmani}, \bibinfo{person}{Jiqun Liu}, {and} \bibinfo{person}{Emine Yilmaz}.}
  \bibinfo{year}{2023}\natexlab{}.
\newblock \showarticletitle{Improving Conversational Recommendation Systems via
  Bias Analysis and Language-Model-Enhanced Data Augmentation}.
\newblock \bibinfo{journal}{\emph{arXiv preprint arXiv:2310.16738}}
  (\bibinfo{year}{2023}).
\newblock


\bibitem[Wen et~al\mbox{.}(2024)]%
        {wen_elaborative_2024}
\bibfield{author}{\bibinfo{person}{Qianfeng Wen}, \bibinfo{person}{Yifan Liu},
  \bibinfo{person}{Joshua Zhang}, \bibinfo{person}{George Saad},
  \bibinfo{person}{Anton Korikov}, \bibinfo{person}{Yury Sambale}, {and}
  \bibinfo{person}{Scott Sanner}.} \bibinfo{year}{2024}\natexlab{}.
\newblock \bibinfo{title}{Elaborative {Subtopic} {Query} {Reformulation} for
  {Broad} and {Indirect} {Queries} in {Travel} {Destination} {Recommendation}}.
\newblock
\urldef\tempurl%
\url{http://arxiv.org/abs/2410.01598}
\showURL{%
\tempurl}
\newblock
\shownote{arXiv:2410.01598 [cs]}.


\bibitem[Williams et~al\mbox{.}(2018)]%
        {williams2018mnli}
\bibfield{author}{\bibinfo{person}{Adina Williams}, \bibinfo{person}{Nikita
  Nangia}, {and} \bibinfo{person}{Samuel Bowman}.}
  \bibinfo{year}{2018}\natexlab{}.
\newblock \showarticletitle{A Broad-Coverage Challenge Corpus for Sentence
  Understanding through Inference}. In \bibinfo{booktitle}{\emph{Proceedings of
  the 2018 Conference of the North American Chapter of the Association for
  Computational Linguistics: Human Language Technologies, Volume 1 (Long
  Papers)}} (New Orleans, Louisiana). \bibinfo{publisher}{Association for
  Computational Linguistics}, \bibinfo{pages}{1112--1122}.
\newblock
\urldef\tempurl%
\url{http://aclweb.org/anthology/N18-1101}
\showURL{%
\tempurl}


\bibitem[Xie et~al\mbox{.}(2024)]%
        {xie2024travelplanner}
\bibfield{author}{\bibinfo{person}{Jian Xie}, \bibinfo{person}{Kai Zhang},
  \bibinfo{person}{Jiangjie Chen}, \bibinfo{person}{Tinghui Zhu},
  \bibinfo{person}{Renze Lou}, \bibinfo{person}{Yuandong Tian},
  \bibinfo{person}{Yanghua Xiao}, {and} \bibinfo{person}{Yu Su}.}
  \bibinfo{year}{2024}\natexlab{}.
\newblock \showarticletitle{Travelplanner: A benchmark for real-world planning
  with language agents}.
\newblock \bibinfo{journal}{\emph{arXiv preprint arXiv:2402.01622}}
  (\bibinfo{year}{2024}).
\newblock


\bibitem[Yin et~al\mbox{.}(2019)]%
        {yin-etal-2019-benchmarking}
\bibfield{author}{\bibinfo{person}{Wenpeng Yin}, \bibinfo{person}{Jamaal Hay},
  {and} \bibinfo{person}{Dan Roth}.} \bibinfo{year}{2019}\natexlab{}.
\newblock \showarticletitle{Benchmarking Zero-shot Text Classification:
  Datasets, Evaluation and Entailment Approach}. In
  \bibinfo{booktitle}{\emph{Proceedings of the 2019 Conference on Empirical
  Methods in Natural Language Processing and the 9th International Joint
  Conference on Natural Language Processing (EMNLP-IJCNLP)}},
  \bibfield{editor}{\bibinfo{person}{Kentaro Inui}, \bibinfo{person}{Jing
  Jiang}, \bibinfo{person}{Vincent Ng}, {and} \bibinfo{person}{Xiaojun Wan}}
  (Eds.). \bibinfo{publisher}{Association for Computational Linguistics},
  \bibinfo{address}{Hong Kong, China}, \bibinfo{pages}{3914--3923}.
\newblock
\href{https://doi.org/10.18653/v1/D19-1404}{doi:\nolinkurl{10.18653/v1/D19-1404}}


\bibitem[Zhang et~al\mbox{.}(2023)]%
        {zhang_recipe-mpr_2023}
\bibfield{author}{\bibinfo{person}{Haochen Zhang}, \bibinfo{person}{Anton
  Korikov}, \bibinfo{person}{Parsa Farinneya}, \bibinfo{person}{Mohammad~Mahdi
  Abdollah~Pour}, \bibinfo{person}{Manasa Bharadwaj}, \bibinfo{person}{Ali
  Pesaranghader}, \bibinfo{person}{Xi~Yu Huang}, \bibinfo{person}{Yi~Xin Lok},
  \bibinfo{person}{Zhaoqi Wang}, \bibinfo{person}{Nathan Jones}, {and}
  \bibinfo{person}{Scott Sanner}.} \bibinfo{year}{2023}\natexlab{}.
\newblock \showarticletitle{Recipe-{MPR}: {A} {Test} {Collection} for
  {Evaluating} {Multi}-aspect {Preference}-based {Natural} {Language}
  {Retrieval}}. In \bibinfo{booktitle}{\emph{Proceedings of the 46th
  {International} {ACM} {SIGIR} {Conference} on {Research} and {Development} in
  {Information} {Retrieval}}} \emph{(\bibinfo{series}{{SIGIR} '23})}.
  \bibinfo{publisher}{Association for Computing Machinery},
  \bibinfo{address}{New York, NY, USA}, \bibinfo{pages}{2744--2753}.
\newblock
\showISBNx{978-1-4503-9408-6}
\href{https://doi.org/10.1145/3539618.3591880}{doi:\nolinkurl{10.1145/3539618.3591880}}


\bibitem[Zheng et~al\mbox{.}(2023)]%
        {zheng2023judging}
\bibfield{author}{\bibinfo{person}{Lianmin Zheng}, \bibinfo{person}{Wei-Lin
  Chiang}, \bibinfo{person}{Ying Sheng}, \bibinfo{person}{Siyuan Zhuang},
  \bibinfo{person}{Zhanghao Wu}, \bibinfo{person}{Yonghao Zhuang},
  \bibinfo{person}{Zi Lin}, \bibinfo{person}{Zhuohan Li},
  \bibinfo{person}{Dacheng Li}, \bibinfo{person}{Eric Xing}, {et~al\mbox{.}}}
  \bibinfo{year}{2023}\natexlab{}.
\newblock \showarticletitle{Judging llm-as-a-judge with mt-bench and chatbot
  arena}.
\newblock \bibinfo{journal}{\emph{Advances in Neural Information Processing
  Systems}}  \bibinfo{volume}{36} (\bibinfo{year}{2023}),
  \bibinfo{pages}{46595--46623}.
\newblock


\bibitem[Zhu et~al\mbox{.}(2024)]%
        {zhu2024llm}
\bibfield{author}{\bibinfo{person}{Lixi Zhu}, \bibinfo{person}{Xiaowen Huang},
  {and} \bibinfo{person}{Jitao Sang}.} \bibinfo{year}{2024}\natexlab{}.
\newblock \showarticletitle{A LLM-based Controllable, Scalable, Human-Involved
  User Simulator Framework for Conversational Recommender Systems}.
\newblock \bibinfo{journal}{\emph{arXiv preprint arXiv:2405.08035}}
  (\bibinfo{year}{2024}).
\newblock


\bibitem[Zhu et~al\mbox{.}(2018)]%
        {zhu2018texygen}
\bibfield{author}{\bibinfo{person}{Yaoming Zhu}, \bibinfo{person}{Sidi Lu},
  \bibinfo{person}{Lei Zheng}, \bibinfo{person}{Jiaxian Guo},
  \bibinfo{person}{Weinan Zhang}, \bibinfo{person}{Jun Wang}, {and}
  \bibinfo{person}{Yong Yu}.} \bibinfo{year}{2018}\natexlab{}.
\newblock \showarticletitle{Texygen: A Benchmarking Platform for Text
  Generation Models}. In \bibinfo{booktitle}{\emph{The 41st International ACM
  SIGIR Conference on Research \& Development in Information Retrieval}} (Ann
  Arbor, MI, USA) \emph{(\bibinfo{series}{SIGIR '18})}.
  \bibinfo{publisher}{Association for Computing Machinery},
  \bibinfo{address}{New York, NY, USA}, \bibinfo{pages}{1097–1100}.
\newblock
\showISBNx{9781450356572}
\href{https://doi.org/10.1145/3209978.3210080}{doi:\nolinkurl{10.1145/3209978.3210080}}


\end{thebibliography}
